\documentstyle[12pt]{article}

\begin{document}

\topmargin -12mm
\oddsidemargin 5mm

\setcounter{page}{1}

\vspace{2cm}
\begin{center}

{\bf USE OF CRYSTALS FOR HIGH ENERGY PHOTON BEAM LINEAR POLARIZATION 
CONVERSION INTO CIRCULAR}\\
\vspace{3mm}
{\large  N.Z. Akopov{$^a$},
A.B. Apyan{$^b$}, S.M. Darbinyan{$^c$}}\\
\vspace{3mm}

{\em  Yerevan Physics Institute, Yerevan, 375036, Armenia}\\
\vspace{3mm}
$^a$ E-mail:akopov@inix.yerphi.am\\
$^b$ E-mail:aapyan@jerewan1.yerphi.am\\
$^c$ E-mail:simon@inix.yerphi.am\\
\end{center}

\vspace{3mm}
\begin{quote}
~~~~The possibility to convert the photon beam linear polarization into 
circular one at photon energies of hundreds GeV with the use of crystals 
is considered. The energy and orientation dependencies of refractive 
indexes are investigated in case of diamond, silicon and germanium 
crystal targets. To maximize the values for figure of merit, the 
corresponding crystal optimal orientation angles and thickness are found.
The degree of circular polarization and intensity of photon beam are 
estimated and possibility of experimental realization is discussed.
\end{quote}

\indent
The method to convert the linear polarization into circular one at high
energies using the single crystals (similar to the quarter-wave plate 
in optics) was proposed by N. Cabibbo [1] since sixties and its
experimental verification is not yet found. In the last decade the
interest to this problem has increased [2] and there have been proposals
to verify it experimentally [3]. This interest is connected with planned
experiments with circular polarized photon beams at energies of
tens and hundreds GeV on investigation of fundamental problems of theory.
It was considered the possibility of using the circularly polarized photon
beams in order to measure G, the polarized gluon distribution in
nucleons, which is necessary for understanding the spin crisis  problem.
The investigation of this problem is of great importance since the results
of EMC Collaboration experiment show that only $30\%$ of nucleon spin is
carried out by quarks [4]. The problem of spin crisis can be investigated 
by the processes of production of jets and heavy quarks via photon-gluon
fusion [5,6,7], production of high transverse momentum mesons [8],
production of $J/\psi$ [9] and $\rho^{0}$-mesons by circularly polarized
photons. In particular for processes of polarized photon fusion with 
gluons of polarized nucleon target the asymmetry in produced
charm-anticharm
quark pairs rates for opposite polarization of the target will appreciate
the gluon contribution to the proton spin. According to theoretical
calculations and existing proposals [6,10] a large asymmetry, of $\sim
40\%$, is anticipated. The estimated value of $\Delta G/G$ obtained
recently in the HERMES experiment (HERA, DESY) at the photon energy of
27.5 GeV is in order 0.41 which corresponds to asymmetry in order of 0.28 
[11].
  
\indent
Circular polarized photon beams can be produced by the longitudinal 
polarized electron beams, however the energy of electrons at modern 
accelerators is not sufficiently high to realize of above mentioned
experiments. Therefore it is nessesary to continue the works on 
production of circular polarized photon beams at high energy proton 
accelerators (CERN, Fermilab). The first stage of the experiment NA59 
devoted to production of linear polarized photon beam was performed at
CERN on the 180 GeV energy SPS electron beam, using the 1.5 cm silicon
radiator. The $40\%$ of the average linear polarization for the photons
at energy range of 90-140 GeV was obtained. The second stage of the
experiment on conversion of the photon beam linear polarization is 
planned to realize  in this year. In this connection is actual to 
carry out calculations on investigation of the problem in different
crystals to choose more convenient one for  polarization conversion
experiment and to estimate the crystal plate optimal thickness.
 
\indent
In present work the problem of photon polarization conversion at 
energies of 100-300 GeV for usually used C, Si and Ge crystals is 
investigated. The energy and orientation dependencies of real parts of 
refractive indexes are calculated as well as the optimal thickness' of
crystal plates, which provide the production of photon beams with 
maximal value of figure of merit (FOM) in sense of the degree of 
circular polarization and beam intensity, are estimated.

\indent
The photon beam orientation with respect to the crystal axes is 
defined in the following way. Let the chosen three orthogonal axes of 
cubic crystal are (1,2,3). The beam orientation is defined by angle 
$\theta$ between the axis 3 and the direction of incident photon beam 
$\vec{n}$ and by the angle $\alpha$ between the projection of $\vec{n}$
on the plane (1,2) and the axis 1. The photon beam polarization direction 
is defined with respect to the incidence plane containing photon
direction $\vec{n}$ and axis 3 and the indexes $\|$ and $\perp$
correspond to cases of $\varphi_{0}$ =0 and $\varphi_{0}= \pi/2$ (Fig.1).
The coordinate system (x, y, z) connected with beam direction is chosen as 
shown in Fig.1. The projections $(g_{x},~g_{y},~g_{z})$ of reciprocal 
lattice vector $\vec{g}=\vec{g}_{1}\vec{n}_{1}+\vec{g}_{2}\vec{n}_{2}+
\vec{g}_{3}\vec{n}_{3}$ are equal:

\begin{eqnarray}
g_{\|}=g_{z}=\theta(G_{1}\cos\alpha+G_{2}\sin\alpha)~, \nonumber\\
g_{x}=G_{1}\cos\alpha+G_{2}\sin\alpha~,\\
g_{y}=-G_{1}\sin\alpha+G_{2}\cos\alpha~, 
\nonumber
\end{eqnarray}
where $\vec{g_{i}}$ are the vectors along the crystal basis axes 
$(|\vec{g_{i}}|=2\pi/\alpha)$ and $G_{1}, G_{2}, G_{3}$ are projections
of $\vec{g}$ on the axes 1, 2, 3.

\indent
When high energy photon beam propagates through medium the photons are 
absorbed mainly due to the pair production mechanism on the medium atoms
and photon beam attenuates with penetrating depth. Hence the 
refractive index is a complex quantity and his imaginary part is 
connected with pair production cross section by equation: Im$ n=W/2\omega$
where $W=N\sigma$ is the absorption coefficient, n is 
dencity of atoms and $\sigma$ is pair production cross section. In 
crystals the cross section (and accordingly refractive index) depends 
upon photon beam polarization direction with respect to the crystal 
axes. The real parts of refractive indexes defined via imaginary parts 
by dispersion relations [1]. This interesting assumption 
for photon birefringence at high energies is not so obviousely and 
its experimental verification presents itself of great importance.

\indent
Let us consider the linear polarized photon beam incidents upon crystal 
as shown in Fig.1. The polarization of photon beam is described by 
Stocks parameters $\xi_{1}=P_{0}\sin 2\varphi_{0},~
\xi_{3}=P_{0}\cos 2\varphi_{0}$, where $P_{0}$ and $\varphi_{0}$ are the 
degree and direction of polarization. The beam intensity and Stocks 
parameters beyond the crystal plate of thickness $l$ are defined by 
formulae [12]:

\begin{eqnarray}
I(l)=I(0)[\cosh (al)+P_{0}\cos (2\varphi_{0}) \sinh (al)]
e^{-\bar {W}l}~, \nonumber\\
\xi_{1}(l)=\frac{P_{0}\sin (2\varphi_{0})\cos (bl)}
{\cosh (al)+P_{0}\cos (2\varphi_{0}) \sinh (al)}~, \nonumber\\
\xi_{2}(l)=\frac{P_{0}\sin (2\varphi_{0})\sin (bl)}
{\cosh (al)+P_{0}\cos (2\varphi_{0}) \sinh (al) }~,\\
\xi_{3}(l)=\frac{P_{0}\cos (2\varphi_{0})\cosh (al)+\sinh (al)}
{\cosh (al)+P_{0}\cos (2\varphi_{0}) \sinh (al)}~, \nonumber
\end{eqnarray}
where $W=(W_{\|}+W_{\bot})/2, ~a=(W_{\|}-W_{\bot})/2,~
b=\omega Re(n_{\bot}-n_{\|})$.

\indent
As it's follows from (1)

\begin{equation}
\label{AA}
\xi^{2}_{1}(l)+\xi^{2}_{2}(l)+\xi^{2}_{3}(l)=1+\frac{P^{2}_{0}-1}
{\cosh (al)+P_{0}\cos (2\varphi_{0}) \sinh (al)}
\end{equation}
and in case of completely linearly polarized incident photon beam 
there is conservation of polarization:
\begin{equation}
\label{AB}
\xi^{2}_{1}(l)+\xi^{2}_{2}(l)+\xi^{2}_{3}(l)=
\xi^{2}_{1}(0)+\xi^{2}_{3}(0)
\end{equation}
and this condition can be used for determination of circular
polarization afther measuring the linear polarization of survived
photon beam.

\indent
The formulae (2) permits to calculate the parameters of surviving photon
beam in dependence of orientation angle $\theta$ and incident beam 
polarization direction $\varphi_{0}$. From expressions for $\xi_{2}(l)$
and $I(l)$ it is clear that crystal must be oriented at angles
corresponding to maximum values of real parts differences in order to
decrease the beam attenuation or the thickness of crystal plate. The
results of calculations for C, Si and Ge crystals at energies 100, 200 and
300 GeV are shown in Fig. 1-3. The photon beam is oriented parallel to the
plane $(1\bar{1}0)$ and makes angle $\theta$ with axis [110]. The
absorption coefficients and, respectively, differences of real parts are
calculated by coherent theory [13], the validity of which is not
questionable since the results
obtained by this theory are in good agreement with experimental data at 
considered angles [12].

\indent
The criterion of crystal efficiency is the maximal value of the figure 
of merit (FOM) FOM=$\xi^{2}_{2}(l)I(l)$ in dependence of orientation angle 
$\theta$ and crystal thickness $l$. For determination of $\theta^{opt}$
and $l^{opt}$ the calculations are carried out at $\varphi_{0}=45^{0}$. 
Such choice of incident beam polarization direction is due to the fact,
that at considered energies the condition $al \ll 1$ is fulfilled and 
$\xi_{2}(l)$ receives its maximum value at that angle. The results 
given in Table 1 show, that optimal angles $\theta^{opt}$ decrease 
with increasing of photon energy being significantly higher than 
characteristic angle of quasiclassical theory at considered energies [12].
The value of FOM is greatest for diamond but silicon crystal is more suitable
for conversion experiment because of the silicon can be grown of required
thickness and does not need cooling as germanium crystal.

\indent
The authors would like to thank Prof. M.L. Ter-Mikaelian for useful 
discussions. This work was supported by the ISTC Grant A-099.

\newpage
\centerline{{\bf{Figure Captions}}}

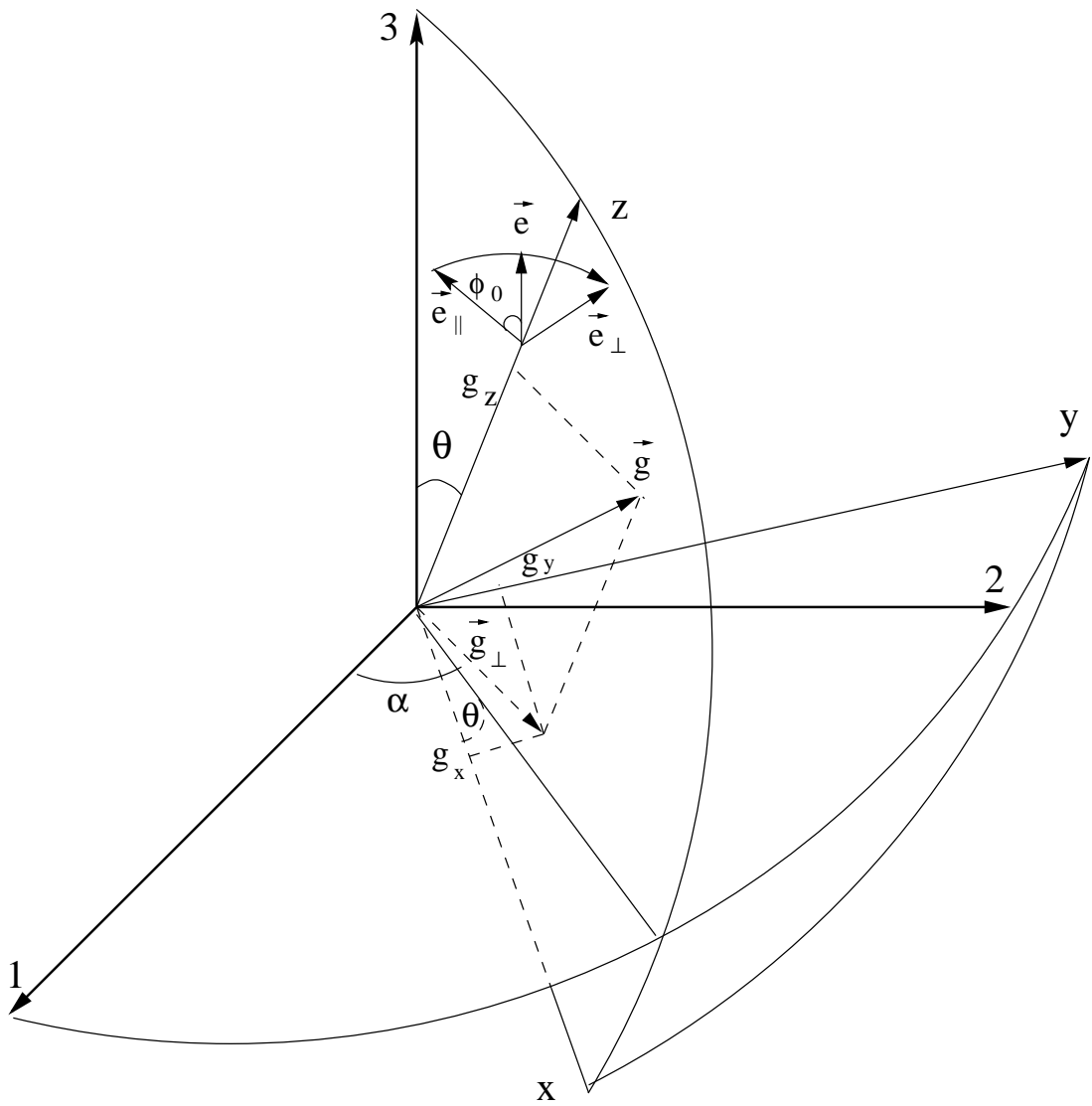
Fig.1. Crystal orientation with respect to the polarized photon beam.

\indent
Fig.2. Curves for diamond, silicon and cooled germanium crystals for 
photon beam
energy 100 GeV. Solid curves for Si, dashed curves for Ge 
and dotted curves for C.\\
a - the differences of refractive indexes real parts as a function 
of $\theta$.\\
b and c - photon beam circular polarization degree and 
intensity as a functions of crystal plate thickness respectively. 

\indent
Fig.3. The curves as in Fig. 2. for photon beam energy 200 GeV.

\indent
Fig.4. The curves as in Fig. 2. for photon beam energy 300 GeV. 
\newpage
\begin{table}[h!]
\begin{center}
\rightline{\bf{Table 1.~~~~~~~~~~~}}
\vspace{2mm}
\begin{tabular}{|l|l|l|l|l|l|l|}

\hline  
Crystal & ~~~$\omega$ & FOM & ~~~$\theta ^{opt}$ & ~~~$l^{opt}$ &
$I(l^{opt})/I(0)$ & $\xi_{2}(l^{opt})$\\
& (GeV) && (mrad) & ~~(cm) && \\
\hline
     & ~~100 & ~3.38~ &~~1.50 &~~5.32~~ &~~~~0.21 &~~0.73\\
~~~C & ~~200 & ~3.56~ &~~0.75 &~~2.76~~ &~~~~0.22 &~~0.75\\
     & ~~300 & ~3.62~ &~~0.50 &~~1.86~~ &~~~~0.23 &~~0.76\\
\hline  
      & ~~100 & ~2.21~ &~~2.29 &~~9.43~~ &~~~~0.17 &~~0.54\\
~~~Si & ~~200 & ~2.63~ &~~1.14 &~~5.50~~ &~~~~0.18 &~~0.61\\
      & ~~300 & ~2.79~ &~~0.76 &~~3.84~~ &~~~~0.19 &~~0.64\\
\hline
        ~~~Ge& ~~100 & ~1.83~ &~~2.38 &~~2.25~~ &~~~~0.16 &~~0.46\\
$(100^{o}K)$ & ~~200 & ~2.18~ &~~1.19 &~~1.32~~ &~~~~0.17 &~~0.53\\
             & ~~300 & ~2.32~ &~~0.79 &~~0.93~~ &~~~~0.17 &~~0.56\\
\hline

\end{tabular}
\end{center}
\label{tab1}
\end{table}
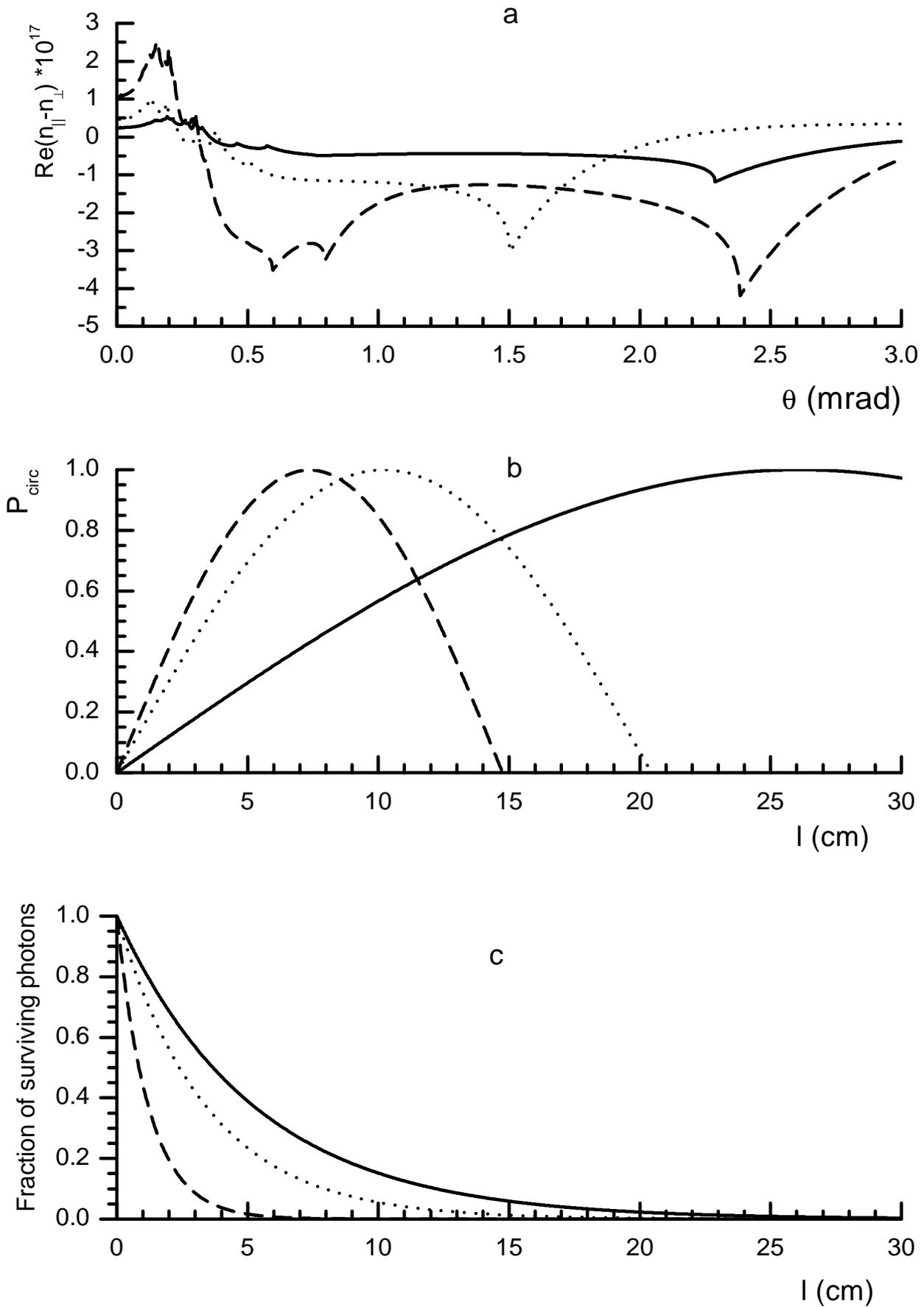
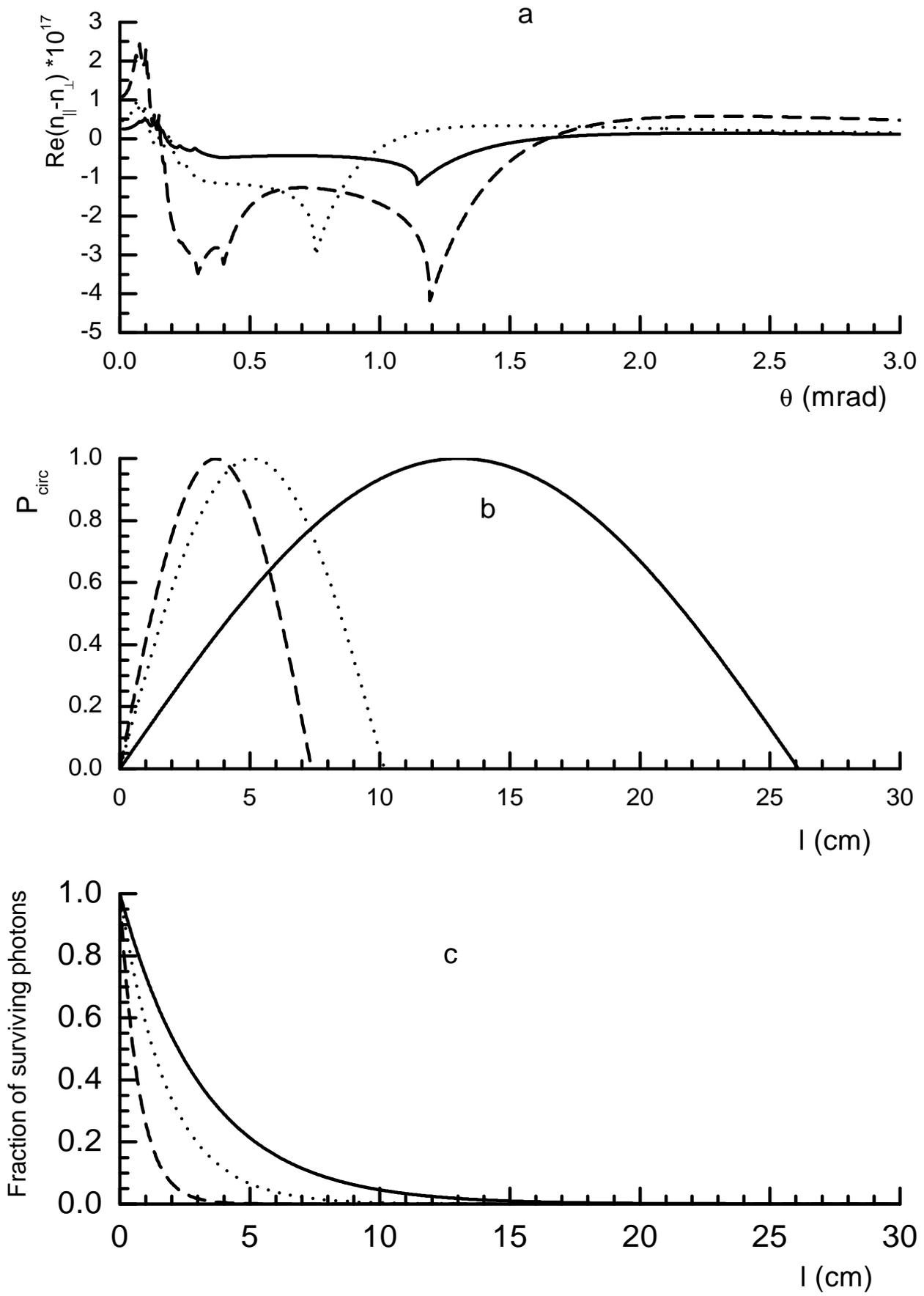
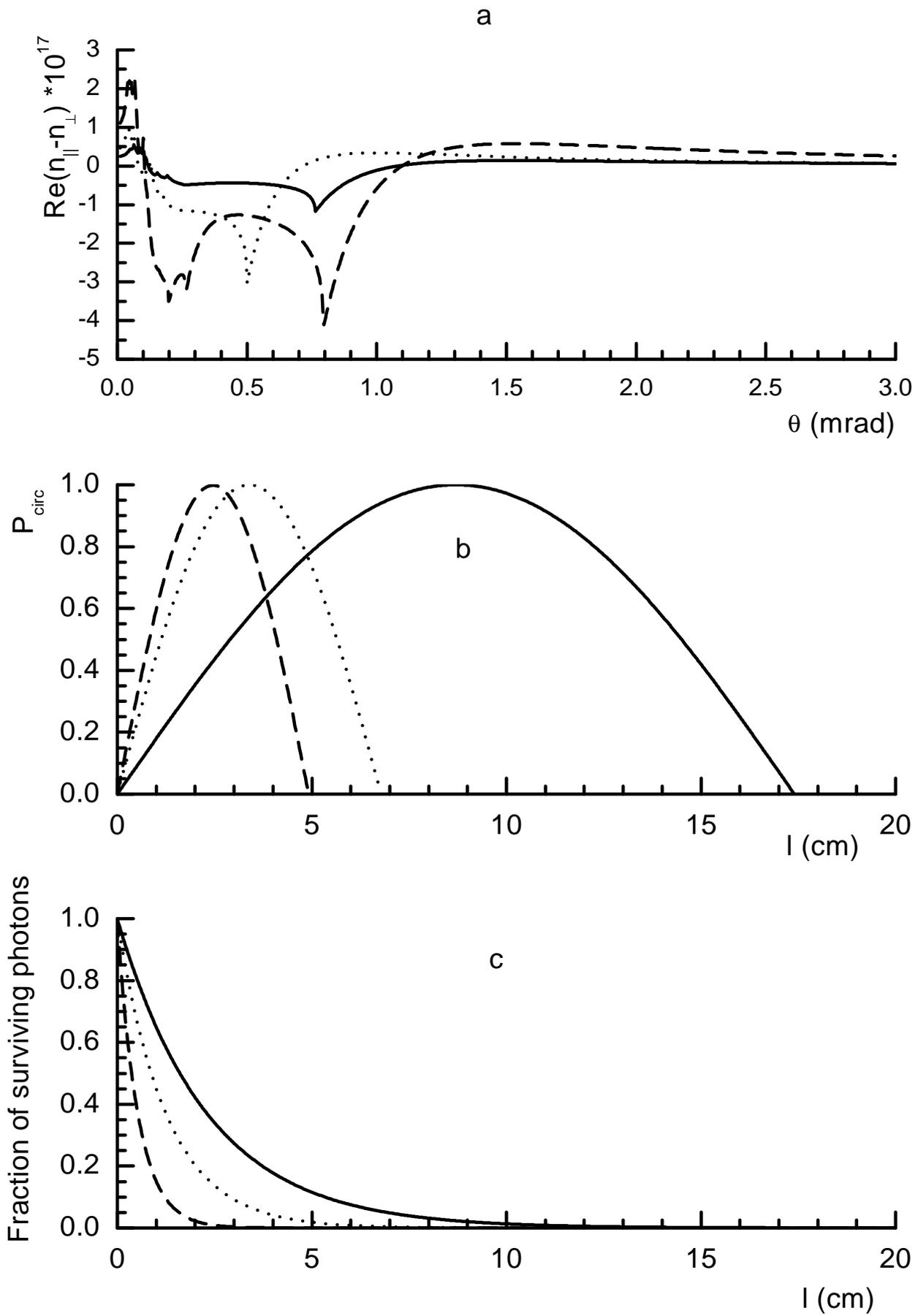
\end{document}